\pdfoutput=1
\documentclass[aps,prd,10pt,twocolumn,superscriptaddress,nofootinbib,floatfix]{revtex4}

\usepackage{mathrsfs, amssymb, amsmath, amsfonts, txfonts, latexsym, graphicx}  
\usepackage[utf8]{inputenc}
	\usepackage[
      colorlinks=true,
      linkcolor=blue,
      urlcolor=cyan,
      filecolor=magenta,
      citecolor=red,
      pdfstartview=FitV,
      hyperfootnotes=false,
      pdftitle={Carrollian c-functions and flat space holographic RG flows in BMS3/CCFT2},
        pdfauthor={Daniel Grumiller, Max Riegler},
        pdfsubject={Carrollian c-functions and flat space holographic RG flows in BMS3/CCFT2},
        pdfkeywords={Carrollian c-functions and flat space holographic RG flows in BMS3/CCFT2},
        bookmarksopen=true
      ]{hyperref}

\newtheorem{thm}{Theorem}

\newcommand{\eq}[2]{\begin{equation} #1 \label{#2} \end{equation}}
\DeclareMathOperator{\extdm}{d}
\newcommand{\extd}{\extdm \!}
\newcommand{\eps}{\varepsilon}

\newcommand{\GN}{G_N}
\providecommand{\Lt}{{\tt L}}
\renewcommand{\Lt}{{\tt L}}
\providecommand{\Mt}{{\tt M}}
\renewcommand{\Mt}{{\tt M}}

\DeclareFontFamily{OT1}{rsfs}{}
\DeclareFontShape{OT1}{rsfs}{m}{n}{ <-7> rsfs5 <7-10> rsfs7 <10->rsfs10}{} 
\DeclareMathAlphabet{\mycal}{OT1}{rsfs}{m}{n}

\begin{document}

%%%%%%%%%%%%%%%%%%%%%%
%%% TITLE/ABSTRACT %%%
%%%%%%%%%%%%%%%%%%%%%%

\title{Carrollian $\boldsymbol{c}$-functions and flat space holographic RG flows in BMS$\boldsymbol{_3}$/CCFT$\boldsymbol{_2}$}

\author{Daniel Grumiller}
\affiliation{Institute for Theoretical Physics, TU Wien, Wiedner Hauptstrasse 8–10/136, A-1040 Vienna,
Austria}
\affiliation{Theoretical Sciences Visiting Program, Okinawa Institute of Science and Technology Graduate University, Onna, 904-0495, Japan}
\email{grumil@hep.itp.tuwien.ac.at}

\author{Max Riegler}
\affiliation{Institute for Theoretical Physics, TU Wien, Wiedner Hauptstrasse 8–10/136, A-1040 Vienna,
Austria}
\affiliation{Quantum Technology Laboratories, Clemens-Holzmeister-Straße 6/6, 1100 Austria, Vienna}
\email{rieglerm@hep.itp.tuwien.ac.at}

\date{\today}

\begin{abstract}
We discuss $c$-functions and their holographic counterpart for two-dimensional field theories with Carrollian conformal fixed points in the UV and the IR. Specifically, we construct asymptotically flat domain wall solutions of three-dimensional Einstein-dilaton gravity that model holographic RG flows between BMS$_3$ invariant UV and IR fixed points. We prove three theorems for such flows: 1.~for every holographic RG flow in AdS$_3$, there is a corresponding one in flat space, 2.~the BMS central charge in the UV cannot be smaller than in the IR, and 3.~the UV/IR ratio of Virasoro central charges is identical to the UV/IR ratio of corresponding BMS central charges. Finally, we tentatively propose a Casini--Huerta-like $c$-functions for BMS$_3$-invariant quantum field theories, inspired by the AdS$_3$/CFT$_2$ relation between monotonicity of the $c$-function and the quantum null energy condition.
\end{abstract}

\maketitle
\tableofcontents

%%%%%%%%%%%%%%%%%%%%%%
%%% BODY OF LETTER %%%
%%%%%%%%%%%%%%%%%%%%%%

\section{Introduction}

Quantum field theories (QFT) adequately describe many systems in nature. A prototypical scenario is a relativistic QFT with conformal fixed points in the ultraviolet (UV) and the infrared (IR). The renormalization group (RG) flow connecting these fixed points can be characterized by $c$-functions that obey a $c$-theorem.

The latter guarantees the existence of some (positive) function, $c(g_i,\,\mu)$, depending on the coupling constants $g_i$ and the RG scale $\mu$ with two key properties: 1.~it decreases monotonically under RG flow towards the IR, and 2.~at the UV and IR fixed points the $c$-function is a (finite) constant. This mathematical statement captures the physical intuition that QFTs have more degrees of freedom in the UV than in the IR.

In two-dimensional (2d) QFTs, Zamolodchikov proved a $c$-theorem by explicitly constructing a $c$-function composed of the energy-momentum tensor components \cite{Zamolodchikov:1986gt}. At the UV- and IR-fixed points, the value of this $c$-function coincides with the respective values of the central charge that characterize the corresponding fixed point conformal field theories (CFT), $c^{\textrm{\tiny UV}}$ and $c^{\textrm{\tiny IR}}$. The $c$-theorem implies $c^{\textrm{\tiny UV}}\geq c^{\textrm{\tiny IR}}$. The $c$-function monotonically interpolates between these fixed point values.

After the advent of AdS/CFT \cite{Maldacena:1997re}, it was natural to seek a holographic version of RG flows and construct holographic versions of $c$-functions. Domain wall solutions in AdS (reviewed below) provide a simple geometric $c$-function \cite{Freedman:1999gp}. Alternatively, Casini and Huerta (CH) proposed a $c$-function \cite{Casini:2004bw} based on entanglement entropy (EE) and one of its indispensable properties, strong subadditivity. The CH $c$-function is tailor-made for AdS/CFT since EE is generally hard to compute in QFTs but simple to compute on the gravity side in terms of minimal \cite{Ryu:2006bv} or extremal \cite{Hubeny:2007xt} surfaces.

The main purpose of our work is to (holographically) construct $c$-functions for 2d field theories with conformal Carrollian fixed points in the UV and IR. The primary motivation for pursuing this goal is a desire for a better understanding of flat space holography and Carrollian CFTs. The main tools we shall employ are flat space domain walls (which we construct and discuss in detail) and flat space holographic EE \cite{Bagchi:2014iea,Basu:2015evh}.

This paper is organized as follows. In section \ref{sec:2}, we review AdS$_3$/CFT$_2$-aspects pertinent to (holographic) $c$-functions. In section \ref{sec:3}, we summarize BMS$_3$/CCFT$_2$ results required for our constructions. In section \ref{sec:4}, we construct domain walls in flat space, discuss their geometric properties, and propose a flat space holographic $c$-function. In section \ref{sec:5}, we prove the three theorems mentioned in the abstract. Section \ref{sec:6} concludes with a tentative CH-inspired proposal for a Carrollian $c$-function.

\section{AdS\texorpdfstring{$\boldsymbol{_3}$}{3}/CFT\texorpdfstring{$\boldsymbol{_2}$}{2} review}\label{sec:2}

This section reviews AdS$_3$/CFT$_2$-aspects pertinent to (holographic) $c$-functions.

In section \ref{sec:2.1}, we provide the definition and main properties of the CH $c$-function. In section \ref{sec:2.2}, we recall the relation to the 2d quantum null energy condition (QNEC$_2$). Section \ref{sec:2.3} summarizes a specific class of domain wall solutions in AdS$_3$ as an example for a holographic model with non-trivial CH $c$-functions. 

\subsection[Casini--Huerta \texorpdfstring{$c$}{c}-function]{Casini--Huerta \texorpdfstring{$\boldsymbol{c}$}{c}-function}\label{sec:2.1}
 
The CH $c$-function \cite{Casini:2004bw, Casini:2006es}
\eq{
c(\ell) = 3\ell\,\frac{\extd S_0}{\extd\ell}
}{eq:CH1}
is constructed from ground state EE $S_0$ and depends on the size $\ell$ of the entangling region. By construction, it is monotonic
\eq{
\frac{\extd c(\ell)}{\extd\ell}\leq 0
}{eq:CH2}
as a consequence of strong subadditivity.

When the inequality \eqref{eq:CH2} is saturated, integrating it twice using \eqref{eq:CH1} yields the result for ground state EE in a CFT$_2$ on the plane \cite{Holzhey:1994we, Calabrese:2004eu},
\eq{
S_0 = \frac{c}{3}\,\ln\frac{\ell}{\eps}
}{eq:CH4}
where $c$ is the UV fixed-point value of $c(\ell)$ and the integration constant $\eps$ is interpreted as UV cutoff.

The limit of vanishing entangling region yields the UV-value of the central charge,
\eq{
c^{\textrm{\tiny UV}} = \lim_{\ell\to 0} c(\ell)\,.
}{eq:CH3}
Similarly, in cases where the theory flows to a CFT$_2$ fixed point in the IR its central charge is obtained in the limit of infinite entangling region.
\eq{
c^{\textrm{\tiny IR}} = \lim_{\ell\to\infty} c(\ell) \leq c^{\textrm{\tiny UV}}
}{eq:CH5}
 
\subsection{Relation to Quantum Null Energy Condition}\label{sec:2.2}

Inserting the definition \eqref{eq:CH1} into the monotonicity condition \eqref{eq:CH2} yields an inequality for up to second derivatives of EE.
\eq{
0\geq  \frac{\extd^2 S_0}{\extd\ell^2}-\frac1\ell\,\frac{\extd S_0}{\extd\ell}+\frac{6}{c(\ell)} \bigg(\frac{\extd S_0}{\extd\ell}\bigg)^2 \geq \frac{\extd^2 S_0}{\extd\ell^2}-\frac1\ell\,\frac{\extd S_0}{\extd\ell}+\frac{6}{c^{\textrm{\tiny UV}}} \bigg(\frac{\extd S_0}{\extd\ell}\bigg)^2 
}{eq:CH17}
The last  expression has an interpretation in terms of variations of EE with respect to null deformations of the interval and can be rewritten as
\eq{
0 \geq \frac{\extd^2 S_0}{\extd\lambda^2} + \frac{6}{c^{\textrm{\tiny UV}}}\, \bigg(\frac{\extd S_0}{\extd\lambda}\bigg)^2 
}{eq:CH18}
where $\lambda$ is the deformation parameter (see section 2.5 in \cite{Ecker:2020gnw}). The combinations of derivatives \eqref{eq:CH18} is the right-hand side of QNEC$_2$ \cite{Bousso:2015mna, Bousso:2015wca, Koeller:2015qmn, Balakrishnan:2017bjg}
\eq{
2\pi\,\langle T\rangle \geq \frac{\extd^2 S}{\extd\lambda^2} + \frac{6}{c^{\textrm{\tiny UV}}} \,\bigg(\frac{\extd S}{\extd\lambda}\bigg)^2 
}{eq:CH19}
for the ground state EE $S_0$, while the left-hand side of QNEC$_2$ contains the expectation value of the null projection of the stress-energy tensor, denoted here by  $T$. Since the Poincar\'e-invariant ground state has $\langle T\rangle=0$, the CH inequality \eqref{eq:CH2} implies QNEC$_2$ for the ground state.

This relation between QNEC$_2$ and the CH $c$-function can guide our proposal for $c$-functions in non-Lorentzian QFTs, provided that quantum energy inequalities are available. In the context of flat space holography, this is indeed the case \cite{Grumiller:2019xna}. 

\subsection[Domain walls in AdS\texorpdfstring{$_3$}{3}]{Domain walls in AdS\texorpdfstring{$\boldsymbol{_3}$}{3}}\label{sec:2.3}

Domain walls are a specific set of geometries describing a holographic RG flow of a QFT$_2$ from a UV to an IR CFT$_2$-fixed point. The geometry dual to such a flow has Poincar\'e invariant slices. In adapted coordinates
\eq{
\extd s^2 = \extd\rho^2 + e^{2A(\rho)}\,\big(-\extd t^2+\extd x^2\big)
}{eq:CH8}
the function $A(\rho)$ characterizes the RG-flow. For any $\rho=\rm const.$ we have Poincar\'e$_2$-invariant slices, i.e., the metric \eqref{eq:CH8} has the Killing vectors $\partial_t$, $\partial_x$ and $x\partial_t+t\partial_x$. Each $\rho=\rho_0=\rm const.$-slice thus induces a 2d flat-space metric $\extd s^2_{(2)}=e^{2A(\rho_0)}\,\big(-\extd t^2+\extd x^2\big)$. There are infinitely many conformal Killing vectors for each such slice, corresponding to the conformal symmetries generated in a CFT$_2$. By convention, the asymptotic region describing the UV is reached in the limit $\rho\to\infty$. 

Domain wall solutions arise, for instance, as solutions to Einstein-dilaton gravity. The bulk action
\eq{
I =\frac{1}{16\pi\GN}\int\extd^3x\sqrt{-g}\,\Big(R-\frac{1}{2}(\partial\phi)^2-V(\phi)\Big)
}{eq:CH6}
with the potential (we use unit AdS-radius)
\eq{ 
V(\phi) = -2 + \frac12\,m^2\phi^2+\dots
}{eq:CH7}
yields the equations of motion
\begin{subequations}
 \label{eq:angelinajolie}
\begin{align}
 R_{\mu\nu}-\frac12\,g_{\mu\nu}\, R &= \frac12\,\partial_\mu\phi\partial_\nu\phi-\frac14\,(\partial\phi)^2g_{\mu\nu}-\frac12\,V(\phi)\,g_{\mu\nu} \\
 \nabla^2\Phi &= \frac{\partial V(\phi)}{\partial\phi}
\end{align}
\end{subequations}
Rewriting the potential in terms of a superpotential $W$
\eq{
V(\phi)=-\frac12\,W(\phi)^2+\frac12\,W'(\phi)^2
}{eq:CH9}
reduces the equations of motion for domain wall solutions \eqref{eq:CH8} to first order equations
\eq{
\frac{\extd A(\rho)}{\extd\rho} = - \frac12\,W(\phi(\rho)) \qquad\qquad \frac{\extd\phi(\rho)}{\extd\rho} = \frac{\extd W(\phi(\rho))}{\extd\phi(\rho)}\,.
}{eq:CH10}

An example is the superpotential
\eq{
W(\phi) = -2 -\frac14\,\phi^2-\frac{\alpha}{8}\,\phi^4
}{eq:CH11}
corresponding to a mass $m^2=-\frac34$. Integrating the equations \eqref{eq:CH10} for this superpotential yields the domain wall solution
\eq{
A(\rho) = \Big(1-\frac{1}{16\alpha}\Big)\,\rho-\frac{j^2}{16(e^\rho-\alpha j^2)}+\frac{\log\big(e^\rho-\alpha j^2\big)}{16\alpha}
}{eq:CH12}
and the scalar field
\eq{
\phi(\rho) = \phi_0 + \frac{j e^{-\rho/2}}{\sqrt{1-\alpha j^2 e^{-\rho}}}
}{eq:CH13}
where $j$ and $\phi_0$ are integration constants. 

For this example, the CH $c$-function was calculated for small $\ell$ \cite{Ecker:2020gnw}
\eq{
c(\ell\ll 1) = c\,\Big(1-\frac{\pi\ell}{64} + {\cal O}(\ell^2)\Big)
}{eq:CH14}
and large $\ell$
\eq{
c(\ell\gg 1) = \frac{c}{1-\frac{1}{16\alpha}} + \dots
}{eq:CH15}
assuming negative $\alpha$. Here $c=c^{\textrm{\tiny UV}}=3/(2\GN)$ takes the Brown--Henneaux value, and we have the relation
\eq{
c^{\textrm{\tiny IR}} =  \frac{c^{\textrm{\tiny UV}}}{1-\frac{1}{16\alpha}} < c^{\textrm{\tiny UV}}\,.
}{eq:CH16}

Alternatively, there is a domain wall holographic $c$-function \cite{Freedman:1999gp}
\eq{
c_{\textrm{\tiny dw}}(\rho) = \frac{c^{\textrm{\tiny UV}}}{A'(\rho)}
}{eq:dwc}
that does not require calculating EE but only uses the (derivative of the) domain wall profile function $A(\rho)$ as input. Since $\lim_{\rho\to\infty}A'(\rho)=1$ and $\lim_{\rho\to-\infty}A'(\rho)=1-\frac{1}{16\alpha}$ we recover the correct UV- and IR-values of the central charge. Moreover, $A'(\rho)$ has the correct monotonicity for a $c$-function.

In our flat space construction below, we shall propose something analogous to the domain wall $c$-function \eqref{eq:dwc}.

\section{BMS\texorpdfstring{$\boldsymbol{_3}$}{3}/CCFT\texorpdfstring{$\boldsymbol{_2}$}{2} summary}\label{sec:3}
 
In this section, we summarize BMS$_3$/CCFT$_2$ results required for our constructions of flat space holographic $c$-functions in later sections.

In section \ref{sec:3.1}, we summarize gravity-aspects of BMS$_3$-invariant QFTs, also known as CCFT$_2$. We collect corresponding field theory aspects in section \ref{sec:3.2}. In section \ref{sec:3.3}, we state the quantum inequalities based on EE that apply to these theories. 

\subsection[Gravity aspects of BMS\texorpdfstring{$_3$}{3}/CCFT\texorpdfstring{$_2$}{2}]{Gravity aspects of BMS\texorpdfstring{$\boldsymbol{_3}$}{3}/CCFT\texorpdfstring{$\boldsymbol{_2}$}{2}}\label{sec:3.1}

We are interested in 2d QFTs invariant under CCFT$_2$ symmetries generated by the $\mathfrak{bms}_3$ algebra 
\begin{subequations}
\label{eq:CH20}
        \begin{align}
            [\Lt_n,\Lt_m] & = (n-m)\,\Lt_{n+m}+\frac{c_{\textrm{L}}}{12}\,n(n^2-1)\,\delta_{n+m,\,0}\\
            [\Lt_n,\Mt_m] & = (n-m)\,\Mt_{n+m}+\frac{c_{\textrm{M}}}{12}\,n(n^2-1)\,\delta_{n+m,\,0}\\
            [\Mt_n,\Mt_m] & = 0\,.
        \end{align}
\end{subequations}        
The generators $\Lt_n$ yield a Virasoro subalgebra with central charge $c_{\textrm{L}}$. They are sometimes referred to as ``superrotations'' in a gravity context. The supertranslation generators $\Mt_n$ produce a central charge $c_{\textrm{M}}$ in the mixed commutator.

The simplest gravity dual leading to \eqref{eq:CH20} as asymptotic symmetries is Einstein gravity with Barnich--Comp\`ere boundary conditions \cite{Barnich:2006av}. 
\begin{multline}
\extd s^2 = {\cal M}(\varphi)\,\extd u^2-2\extd u \extd r + r^2\,\extd\varphi^2 \\
+ \big(2{\cal L}(\varphi)+u{\cal M}^\prime(\varphi)\big)\,\extd u\extd\varphi + \dots 
\label{eq:CH21}
\end{multline}
The coordinate ranges are $u,r\in\mathbb{R}$ and either $\varphi\sim\varphi+2\pi$ or $\varphi\in\mathbb{R}$. The ellipsis denotes subleading terms in a large-$r$ expansion. The state-dependent functions $\cal L,M$ appear as integrands in the boundary charges. Their (Fourier-) modes generate the $\mathfrak{bms}_3$ algebra \eqref{eq:CH20} as asymptotic symmetry algebra, with central charges $c_{\textrm{L}}$, $c_{\textrm{M}}$ the values of which depend on the gravity theory, see e.g.~\cite{Bagchi:2012yk}.

For Einstein gravity without cosmological constant, the Virasoro central charge $c_{\textrm{L}}$ vanishes since there is no dimensionless coupling constant, while the mixed central charge $c_{\textrm{M}}$ is non-zero \cite{Barnich:2006av}. We use a (standard) normalization of the generators $\Mt_n$ where $c_{\textrm{M}}=3/\GN$.  

The null orbifold (${\cal M}={\cal L}=0$) \cite{Cornalba:2002fi}
\eq{
\extd s^2 =  -2\extd u \extd r + r^2\,\extd\varphi^2
}{eq:CH22}
is dual to the ground state of the BMS$_3$ invariant QFT in the same way that Poincar\'e-patch AdS$_3$ is the gravity dual of the ground state of a QFT on the plane with a CFT$_2$ fixed point in the UV. If $\varphi\sim\varphi+2\pi$, the null orbifold has a singularity in the causal structure at $r=0$. For our purposes, this is as irrelevant as the coordinate singularity in the Poincar\'e patch horizon since we will construct domain wall solutions that only asymptote to the null orbifold but do not exhibit its singular behavior in the interior. Moreover, we can simply decompactify $\varphi$.

We later investigate what happens when the retarded time coordinate $u$ gets rescaled by some factor $\lambda$ (absorbing such factors in the state-dependent functions $\cal L,M$ when possible). 
\begin{multline}
\extd s^2 = {\cal M}(\varphi)\,\extd u^2-2\lambda\,\extd u \extd r + r^2\,\extd\varphi^2 \\
+ \big(2{\cal L}(\varphi)+u{\cal M}^\prime(\varphi)\big)\,\extd u\extd\varphi + \dots 
\label{eq:CH21a}
\end{multline}
The superrotation charges $\Lt_n$ are associated with asymptotic Killing vectors $e^{in\varphi}\partial_\varphi$ and hence unaffected by a rescaling of retarded time. By contrast, the supertranslation charges $\Mt_n$ are associated with asymptotic Killing vectors $e^{in\varphi}\partial_u$ and thus get rescaled by $1/\lambda$.
\eq{
[\Lt_n,\,\Mt_m/\lambda] = (n-m)\,\Mt_{n+m}/\lambda + \frac{c_{\textrm{M}}}{12}\,n(n^2-1)\,\delta_{n+m,\,0}
}{eq:CH22a}
Effectively, this rescales the BMS central charge by $\lambda$.
\eq{
[\Lt_n,\,\Mt_m] = (n-m)\,\Mt_{n+m} + \frac{\lambda\, c_{\textrm{M}}}{12}\,n(n^2-1)\,\delta_{n+m,\,0}
}{eq:CH22b}
This observation will be crucial for holographic RG flows modeled by flat space domain walls, discussed in section \ref{sec:4}.

\subsection{Field theory aspects}\label{sec:3.2}

We now summarize some aspects of CCFTs; see \cite{Bagchi:2019xfx, deBoer:2023fnj} and refs.~therein for more details.

We start by recalling the definition of the Carrollian conformal weights, analogous to conformal weights:
\eq{
\Lt_0|h_{\textrm{L}},h_{\textrm{M}}\rangle = h_{\textrm{L}} |h_{\textrm{L}},\,h_{\textrm{M}}\rangle\qquad\;\,\Mt_0|h_{\textrm{L}},h_{\textrm{M}}\rangle = h_{\textrm{M}}|h_{\textrm{L}},h_{\textrm{M}}\rangle
}{eq:CH45a}
While the interpretation of the Virasoro central charge $c_\textrm{L}$ in the $\mathfrak{bms}_3$ algebra \eqref{eq:CH20} is analogous to the corresponding CFT$_2$ interpretation, the interpretation of the mixed central charge $c_\textrm{M}$ is more subtle. At first glance, its precise value seems irrelevant, since a change of basis $\Mt_n\to \lambda \Mt_n$ is an automorphism of the $\mathfrak{bms}_3$ algebra upon rescaling $c_\textrm{M}$. Nevertheless, there is a Cardy-like entropy formula \cite{Barnich:2012xq,Bagchi:2012xr} (see also \cite{Riegler:2014bia,Fareghbal:2014oba}), which for $c_\textrm{L}=0$ (and $h_\textrm{M}>0$) reads
\eq{   
S = 2\pi\, h_\textrm{L}\,\sqrt{\frac{c_\textrm{M}}{24 h_\textrm{M}}}\,.
}{eq:Cardy} 
The reason there is no contradiction between the appearance of $c_\textrm{M}$ in the entropy formula \eqref{eq:Cardy} and the fact that its value can be rescaled to an arbitrary (positive) number is decisive to understanding our RG flow results presented in later sections. The simple point is that whenever $c_\textrm{M}$ appears in dimensionless ratios, there is a meaning to this ratio. In the Cardy-like formula \eqref{eq:Cardy}, the combination $c_\textrm{M}/h_\textrm{M}$ is dimensionless. Therefore, this formula makes sense. (Another way to come to the same conclusions is to note that both $h_\textrm{M}$ and $c_\textrm{M}$ scale in the same way under the automorphism $\Mt_n\to\lambda\,\Mt_n$, so that the entropy \eqref{eq:Cardy} is invariant under it.) The lesson for later is that the value of $c_\textrm{M}$ by itself is physically irrelevant, but dimensionless ratios involving $c_\textrm{M}$ can be physically relevant.

We focus now on the main observable of interest, EE. It was calculated in \cite{Bagchi:2014iea} for the null orbifold, the global flat space vacuum, and for thermal states, and in \cite{Grumiller:2019xna} for any vacuum-like state, including arbitrary BMS$_3$-descendants of the vacuum and of thermal states. We shall need only the result for the null orbifold.
\eq{
S_{\textrm{EE}} = S_{\textrm{L}}+S_{\textrm{M}}
}{eq:CH23}
with 
\eq{
        S_{\textrm{L}} = \frac{c_{\textrm{L}}}{6}\log\frac{\Delta\varphi}{\epsilon_\varphi}\qquad\qquad S_{\textrm{M}} = \frac{c_{\textrm{M}}}{6}\,\bigg(\frac{\Delta u}{\Delta\varphi} - \frac{\epsilon_u}{\epsilon_\varphi}\bigg)
}{eq:CH24}
where $\Delta\varphi$ ($\Delta u$) is the spatial (temporal) extent of the entangling region, $\epsilon_\varphi$, $\epsilon_u$ are UV cut-offs, and $c_{\textrm{L}}$, $c_{\textrm{M}}$ are the central charges of the $\mathfrak{bms}_3$ algebra \eqref{eq:CH20}. The result \eqref{eq:CH24} was confirmed in holographic calculations \cite{Basu:2015evh,Jiang:2017ecm,Apolo:2020bld,Apolo:2020qjm}. For $c_{\textrm{M}}=0$ \eqref{eq:CH24} coincides with one chiral half of ground state EE in a CFT$_2$ \eqref{eq:CH4}. The UV cut-offs $\epsilon_\varphi$, $\epsilon_u$ drop out in the quantum inequalities and our proposal for the $c$-functions discussed below, so we do not discuss them further.

\subsection[Quantum energy conditions in CCFT\texorpdfstring{$_2$}{2}]{Quantum energy conditions in CCFT\texorpdfstring{$\boldsymbol{_2}$}{2}}\label{sec:3.3}

By analogy to AdS$_3$/CFT$_2$, our main interests are quantum energy conditions \cite{Grumiller:2019xna}. We define the expectation values
\eq{
2\pi\langle{\cal T}_M\rangle = \frac{c_{\textrm{M}}}{24}{\cal M}\qquad\; 2\pi\langle{\cal T}_L\rangle = \frac{c_{\textrm{L}}}{24}{\cal M} + \frac{c_{\textrm{M}}}{24}\big(2{\cal L}+u{\cal M}^\prime\big)
}{eq:CH25}
with conventional normalizations (primes denote $\varphi$-de\-ri\-va\-tives). The quantum energy condition for theories with $c_{\textrm{M}}=0$ 
\eq{
2\pi\,\langle{\cal T}_L\rangle \geq S_L''+\frac{6}{c_{\textrm{L}}}\,S^{\prime\,2}_L
}{eq:CH26}
is a chiral half of the QNEC$_2$ inequalities and essentially equivalent to \eqref{eq:CH19}. However, we are more interested in the opposite case, when $c_{\textrm{L}}=0$ but $c_{\textrm{M}}\neq 0$. In that case, the quantum energy condition is (dots denote $u$-derivatives)
\eq{
2\pi\,\langle{\cal T}_M\rangle \geq \dot S_M'+\frac{6}{c_{\textrm{M}}}\,\dot S^2_M\,.
}{eq:CH27}

The quantum inequalities above inspire the CH-like proposal of $c$-functions in CCFT$_2$ at the end of our paper.

%----------------------------------------------
\section{Domain walls in flat space}\label{sec:4}
%----------------------------------------------

In this section, we construct domain wall solutions in 3d flat space, intending to generate holographic RG-flows analogous to the ones discussed in section \ref{sec:2.3}.

In section \ref{sec:4.1}, we set the stage by deriving the possible geometries of flat space domain walls. Section \ref{sec:4.2} focuses on domain wall solutions in flat space Einstein-dilaton gravity. In section \ref{sec:4.3}, we pick specific solutions that allow a holographic RG flow interpretation.
 
\subsection{Geometric aspects of flat space domain walls}\label{sec:4.1}

In AdS$_3$, domain walls are constructed by requiring Poin\-car\'e$_2$ invariance on each slice; alternatively, we could have demanded that the conformal Killing vectors of each 2d slice generate CFT$_2$ symmetries. We follow the second approach to construct flat space domain walls and demand that the degenerate (Carrollian) induced metric has conformal Killing vectors that generate BMS$_3$ symmetries.

Looking at the asymptotic expansion \eqref{eq:CH21}, it is suggestive to consider as ansatz the degenerate (Carrollian) induced metric
\eq{
\extd s^2_{(2)} = g_{\mu\nu}^{(2)}\,\extd x^\mu\extd x^\nu = e^{2A(r)}\,\extd\varphi^2 + 0\cdot\extd u^2 + 0\cdot \extd u\extd\varphi\,.
}{eq:CH31}
This ansatz ensures that even at finite values of the radial coordinate $r$, all slices have the same features as in the asymptotic limit $r\to\infty$. The function $A(r)$ is arbitrary at this stage. Since the induced metric \eqref{eq:CH31} is degenerate, we make sure not to use its inverse in any of our considerations.

To verify this ansatz, we solve the conformal Killing equation
\eq{
\xi^\mu\partial_\mu g_{\alpha\beta}^{(2)} + g_{\alpha\mu}^{(2)}\partial_\beta\xi^\mu + g_{\beta\mu}^{(2)}\partial_\alpha\xi^\mu = g_{\alpha\beta}^{(2)}\,\partial_\mu\xi^\mu
}{eq:CH33}
for the vector field $\xi$ using the degenerate metric \eqref{eq:CH31}. The result
\eq{
\xi = \big(\xi_M(\varphi)+u\,\xi_L^\prime(\varphi)\big)\,\partial_u + \xi_L(\varphi)\,\partial_\varphi
}{eq:CH34}
shows that the conformal Killing vectors \eqref{eq:CH34} indeed generate centerless $\mathfrak{bms}_3$ as Lie-bracket algebra (compare e.g.~with \cite{Oblak:2016eij}). 
\begin{align}
& \big[\xi\big(\xi_M^{(1)},\,\xi_L^{(1)}\big),\,\xi\big(\xi_M^{(2)},\,\xi_L^{(2)}\big)\big]_{\textrm{\tiny Lie}} = 
\label{eq:CH35} \\
& \xi\big(\xi_M^{(1)}\xi_L^{(2)\,\prime}+\xi_L^{(1)}\xi_M^{(2)\,\prime}-\xi_M^{(2)}\xi_L^{(1)\,\prime}-\xi_L^{(2)}\xi_M^{(1)\,\prime},\,\xi_L^{(1)}\xi_L^{(2)\,\prime}-\xi_L^{(2)}\xi_L^{(1)\,\prime}\big) \nonumber
\end{align}
Therefore, $r=\rm const.$ slices in flat space domain walls only contain the term $e^{2A(r)}\,\extd\varphi^2$.

The 3d metric describing flat space domain walls
\eq{
\boxed{\phantom{\Big(}
\extd s^2 = -e^{A(r)}\,2\extd u \extd r + e^{2A(r)}\,\extd\varphi^2
\phantom{\Big)}}
}{eq:CH32}
depends on one arbitrary function\footnote{%
One could add another function $B(r)$ in the first term, but we have eliminated it by fixing the diffeomorphisms $r\to f(r)$ suitably. The gauge choice \eqref{eq:CH32} ensures that both metric coefficients remain bounded and never change sign, provided the function $A(r)$ remains bounded.} 
of the radial coordinate, $A(r)$. The additional assumption implicit in \eqref{eq:CH32} is that we keep Eddington--Finkelstein gauge in the interior of the bulk. This is analogous to keeping Gaussian normal coordinates in the bulk of AdS$_3$ domain walls \eqref{eq:CH8}. 

Let us now address curvature invariants. The Ricci tensor has a single non-zero component.
\eq{
R_{rr} = -A''
}{eq:CH33a}
Regardless of the choice of $A(r)$, all geometries \eqref{eq:CH32} have vanishing scalar curvature invariants and vanishing Cotton tensor. This means these geometries are not only locally conformally flat, but it also implies we need some Page-like curvature invariants \cite{Page:2008yb} if we want to characterize these geometries. 

An example of such an invariant is
\eq{
P = \frac{(R_{\mu\nu}k^\mu k^\nu)^2}{\big(\nabla_\mu\nabla_\nu R_{\alpha\beta}\big)k^\mu k^\nu k^\alpha k^\beta}
}{eq:alt3} 
where $k^\mu$ is any vector with non-vanishing $r$-component.\footnote{%
In Page's construction, $k^\mu$ had to be null, and the scalar invariants were the maximum and minimum with respect to changes of directions of $k^\mu$. In our case, the Ricci tensor is so simple that the quantity $P$ is constant not only under changes of direction but also under changes of the signature of $k^\mu$ from light-like to time-like or space-like; the only requirement that $k^\mu$ has to fulfill is that $R_{\mu\nu}k^\mu k^\nu$ does not vanish unless $R_{\mu\nu}=0$, which in our coordinates implies $k^r\neq 0$.} 
However, in the following subsection, we identify an even simpler and more useful scalar invariant, namely the matter scalar, so we will not employ \eqref{eq:alt3}.

\subsection{Flat space domain walls in Einstein-dilaton gravity}\label{sec:4.2}

Above, we discussed the kinematics of flat space domain walls. Here, we focus on the dynamics of these domain walls. As for AdS$_3$, we consider Einstein-dilaton gravity \eqref{eq:CH6} with field equations \eqref{eq:angelinajolie}. Remarkably, the field equations hold for any choice of the function $A(r)$ in the flat space domain wall \eqref{eq:CH32} provided the scalar field potential vanishes, $V(\phi)=0$, and the scalar field obeys the ordinary differential equation
\eq{
\frac12\,\phi^{\prime\,2} = -A''\,.
}{eq:CH34a}
The geometric reason for this surprising result is that both $(\partial\phi)^2$ and $\nabla^2\phi$ vanish on flat space domain wall backgrounds \eqref{eq:CH32} for any scalar field $\phi$ that is independent of $u$ and $\varphi$.

This result implies that on-shell, the combination appearing in the Ricci tensor \eqref{eq:CH33a} is related to (the derivative of) the scalar field. Thus, we can use the scalar field $\phi$ as a scalar invariant that fully characterizes our geometry. (The additive integration constant contained in $\phi$ does not play any role for geometric properties and can be chosen conveniently; we fix it by demanding $\lim_{r\to\infty}\phi(r)\to 0$.) Moreover, we can either provide the function $A$ as input and determine $\phi$ by integrating once \eqref{eq:CH34a}, or we provide $\phi$ as input and determine $A$ by integrating twice \eqref{eq:CH34a}.

Demanding compatibility with asymptotic flatness requires the expansion
\eq{
A(r\gg 1) = r -r_0 + o(1) 
}{eq:CH35a}
for the remaining function $A(r)$. Introducing the new radial coordinate $\rho=e^{r-r_0}$ leads to the desired asymptotic expansion of the metric
\eq{
\extd s^2 = -2\extd u\extd\rho + \rho^2\,\extd\varphi^2 + \dots
}{eq:CH36}
In the interior the bulk metric 
\eq{
\extd s^2 = -2e^{A(r)}\,\extd u \extd r + e^{2A(r)}\,\extd\varphi^2
}{eq:CH37}
is free from singularities as long as the function $A(r)$ remains finite; in particular, the null orbifold singularity at $r=0$ is absent since the factor $e^{2A(r)}$ always is finite in the interior.

The asymptotic expansion for the scalar field compatible with \eqref{eq:CH35a} follows from integrating the equations of motion \eqref{eq:CH34a}. Since both the leading and the first subleading terms in \eqref{eq:CH35a} drop out in $A''$, only the terms that decay at $r\to\infty$ contribute to the scalar field.
\eq{
\phi(r\gg 1) = \phi_0 + o(1)
}{eq:CH40}
Without loss of generality, we set the integration constant to zero, $\phi_0=0$. For example, if the sub-subleading term in $A$ scales like $e^{-r}$, then the first term in the large-$r$ expansion of $\phi$ decays like $e^{-r/2}$.

We highlight an important subtlety. Reality of the field configuration requires the inequality
\eq{
A'' \leq 0
}{eq:CH41}
in the whole range of definition of the function $A$.  Thus, when designing flat space domain walls by choosing some function $A(r)$, it is crucial to obey the concavity condition \eqref{eq:CH41} for all values of the radial coordinate $r$.

If we slightly change the asymptotic behavior \eqref{eq:CH35a},
\eq{
A(r\gg 1) = \lambda^{-1}\,\big(r -r_0\big) + o(1) \qquad \lambda\in\mathbb{R}^+
}{eq:CH38}
and use the radial coordinate $\rho=e^{(r-r_0)/\lambda}$, the asymptotic expansion of the metric
\eq{
\extd s^2 = -2\lambda\,\extd u\extd\rho + \rho^2\,\extd\varphi^2 + \dots 
}{eq:CH39}
shows that the first term is rescaled by $\lambda$. We will exploit this property in the next subsection to establish holographic RG flows.

\subsection{Flat space holographic RG flow example}\label{sec:4.3} 

We are finally able to model holographic RG flows for BMS$_3$ invariant QFTs. 

One possibility to generate flat space domain walls is to take any AdS$_3$ domain wall for some potential, take the scalar field appearing in that solution as input, and construct the function $A$ by integrating twice \eqref{eq:CH34a}. Any such choice generates a legitimate flat space domain wall; a subset of them generates an associated holographic RG flow between BMS$_3$ invariant UV and IR fixed points. We will be more precise and general about this in section \ref{sec:5}. For now, we focus on a specific example.

\begin{figure}
\begin{center}
 \includegraphics[width=0.7\linewidth]{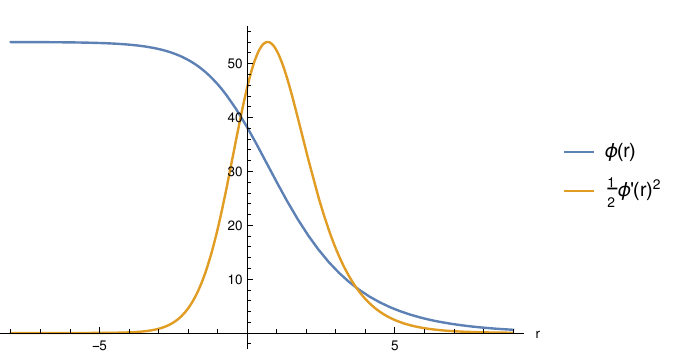}
 \end{center}
 \caption{Scalar field \eqref{eq:CH13} and bulk energy $\frac12\,\phi'(r)^2$ for $\phi_0=0$, $j=54$ and $-\alpha j^2=1$}
 \label{fig:1}
\end{figure}

In our first example, we pick the same scalar field as for AdS$_3$ domain walls \eqref{eq:CH13} (renaming the radial $\rho$ into $r$). In Fig.~\ref{fig:1}, we plot the scalar field $\phi(r)$ and the associated bulk energy $\frac12\,\phi^\prime(r)^2$ for the choices  $\phi_0=0$, $j=54$ and $-\alpha j^2=1$. The blue curve depicts the scalar field and has a clear kink-like structure, interpolating between two different asymptotic values. The orange curve shows that the bulk energy is localized in the interior of the bulk. Its maximum is at $r=\ln 2$ [for general $\alpha, j$ the maximum is at $r=\ln(-2\alpha j^2)$].

Integrating twice \eqref{eq:CH34a} yields
\eq{
A(r) = A_1\, r + A_0 - \frac{j^2}{16}\,\bigg(\frac{1}{e^r - \alpha j^2} + \frac{r-\ln\big(e^r - \alpha j^2\big)}{\alpha j^2}\bigg)
}{eq:CH42}
with two integration constants $A_1$, $A_0$. To obtain the desired asymptotics \eqref{eq:CH35a} we fix $A_1=1$, yielding
\eq{
A(r\to\infty) = r + A_0 - \frac{j^2}{8}\,e^{-r} + {\cal O}(e^{-2r})\,.
}{eq:CH44}

Depending on the sign of $\alpha$, there are different possibilities. In the AdS case, we needed negative $\alpha$ to generate domain walls with a CFT$_2$ fixed point in the IR. We check now whether something analogous is true for the corresponding flat space domain wall.

For negative $\alpha$, there is no singularity in $A(r)$ for any finite value of $r$. Therefore, the coordinate range of this domain wall is $(-\infty,\,\infty)$, and we obtain a second asymptotic region at $r\to-\infty$. In this limit, the function \eqref{eq:CH42} expands as
\eq{
A(r\to-\infty) = \Big(1-\frac{1}{16\alpha}\Big)\,r + A_0 + \frac{1+\ln\big(-\alpha j^2\big)}{16\alpha} + {\cal O}(e^{2r})\,.
}{eq:CH43}

Denoting the UV central charge by $c_\textrm{M}^{\textrm{\tiny UV}}$, comparison of the two asymptotic expansions \eqref{eq:CH44} and \eqref{eq:CH43} yields a result for the IR central charge
\eq{
c_\textrm{M}^{\textrm{\tiny IR}} = \frac{c_\textrm{M}^{\textrm{\tiny UV}}}{1-\frac{1}{16\alpha}} < c_\textrm{M}^{\textrm{\tiny UV}}
}{eq:CH45}
according to the discussion at the end of section \ref{sec:3.1}. Note that the ratio $c_\textrm{M}^{\textrm{\tiny IR}}/c_\textrm{M}^{\textrm{\tiny UV}}\leq 1$ is dimensionless, and hence equation \eqref{eq:CH45} is meaningful [compare with the discussion after the Cardy-like formula \eqref{eq:Cardy}].

\begin{figure}
\begin{center}
 \includegraphics[width=0.7\linewidth]{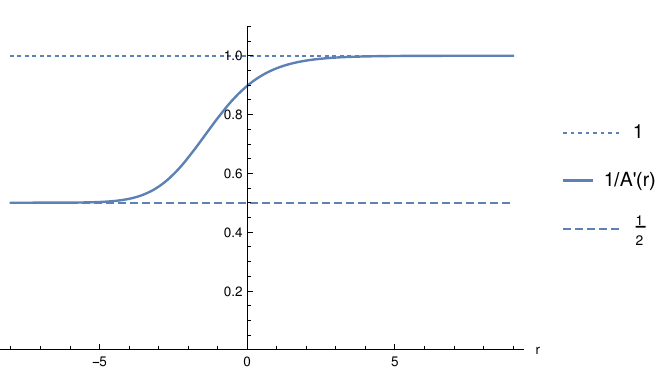}
 \end{center} 
 \caption{Plot of $1/A'(r)$ for $j=1$ and $-\alpha=\frac{1}{16}$ shows it behaves like a $c$-function}
 \label{fig:2} 
\end{figure}

As evident from the plot in Fig.~\ref{fig:2}, the function 
\eq{
c_{\textrm{\tiny dw}}(r) := \frac{c_{\textrm{M}}^{\textrm{\tiny UV}}}{A'(r)}
}{eq:c}
is a $c$-function for this domain wall solution since it approaches the correct UV and IR values and is monotonically decreasing towards the IR.

The result \eqref{eq:CH45} is precisely the same relation as for the corresponding holographic RG flow in AdS$_3$ (with the Virasoro central charge replaced by the $\mathfrak{bms}_3$ central charge $c_\textrm{M}$), see \eqref{eq:dwc}. In the next section, we shall prove that this is not a coincidence but a generic feature relating AdS$_3$ and flat space domain walls and their corresponding RG flow interpretations.

Before generalizing our results, consider the case of positive $\alpha$. In AdS$_3$/CFT$_2$ such ``domain walls'' do not model an RG flow from a UV to an IR fixed point, but rather an RG flow from a UV fixed point to the IR, but without CFT$_2$ fixed point in the IR. As we now show, something comparable happens in flat space. Indeed, for positive $\alpha$ the IR boundary is at finite value of $r$,
\eq{
r_{\textrm{\tiny IR}} = \ln\big(\alpha j^2\big)\,.
}{eq:CH46}
The scalar field and the Ricci tensor are singular at the IR boundary, so in this case, the flow ends at a naked singularity on the gravity side, and there is no BMS$_3$ field theory interpretation in the IR.

\begin{figure}
\begin{center}
 \includegraphics[width=0.7\linewidth]{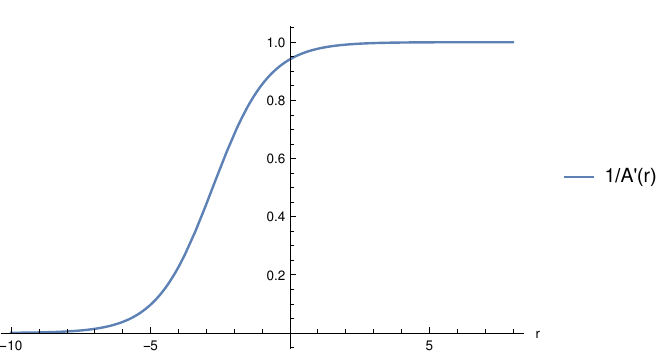}
 \end{center} 
 \caption{Plot of $1/A'(r)$ for $j=1$ and $\alpha=0$ shows the $c$-function vanishes in the IR}
 \label{fig:3} 
\end{figure}

Finally, we consider the limiting case of vanishing $\alpha$. Here the $c$-function tends to zero in the IR (which is again obtained in the limit $r\to-\infty$); see Fig.~\ref{fig:3}. In this case, the IR fixed point is a trivial BMS$_3$-invariant QFT with vanishing central charge $c_\textrm{M}=0$. All these features are analogous to corresponding AdS$_3$/CFT$_2$ features, see e.g.~the discussion in \cite{Ecker:2020gnw}.

\section{Flat space holographic RG flow theorems}\label{sec:5} 

In this section, we state and prove three theorems for RG flows modeled by domain wall solutions in 3d flat space Einstein-dilaton gravity.

In section \ref{sec:5.0}, we collect some definitions used in all theorems. In section \ref{sec:5.1}, we state and prove a correspondence theorem, relating all AdS$_3$ domain walls to corresponding flat space domain walls. In section \ref{sec:5.2}, we state and prove a monotonicity theorem, showing that bulk unitarity implies a monotonically decreasing $c$-function. Putting together both theorems, we prove a third one that shows the equivalence of the UV/IR ratios of Virasoro and $\mathfrak{bms}_3$ central charges. 

\subsection{Definitions}\label{sec:5.0}

In this whole section, we are solely concerned with holographic RG flows generated by flat space domain wall solutions described in section \ref{sec:4}. For these domain walls, we found a $c$-function \eqref{eq:c}. While this definition was based on a single example studied in section \ref{sec:4.3}, it is natural to define generically the {\em flat space holographic $c$-function}
\eq{
\boxed{\phantom{\Big(}
c_{\textrm{\tiny dw}}(r) := \frac{c_{\textrm{M}}^{\textrm{\tiny UV}}}{A'(r)}\qquad\quad \textrm{fixing\;}\lim_{r\to\infty}A(r)=r+{\cal O}(1)\,.
\phantom{\Big)}}
}{eq:cdef}
This formally coincides with the AdS domain wall $c$-function discovered in the seminal work \cite{Freedman:1999gp}.

We define the term {\it proper domain wall solution of AdS$_3$-Einstein-dilaton gravity} to mean an exact solution of the equations of motion \eqref{eq:angelinajolie} with some scalar potential of the form \eqref{eq:CH7} such that in domain wall coordinates \eqref{eq:CH8} metric and scalar field have the following properties:
\begin{enumerate}
 \item for $\rho\to\infty$ the metric asymptotes to Poincar\'e patch AdS$_3$ with unit AdS-radius and the scalar field approaches zero
 \item for $\rho\to-\infty$ the metric asymptotes to Poincar\'e patch AdS$_3$ with AdS-radius smaller than one and the scalar field approaches a constant (that can be zero)
 \item for finite values of $\rho$ the metric function $A(\rho)$ and the scalar field $\phi(\rho)$ are bounded real functions; moreover, $A(\rho)$ is at least $C^2$ and $\phi(\rho)$ at least $C^1$
\end{enumerate}

Similarly, we define the term {\it proper flat-space domain wall solution} to mean an exact solution of the equations of motion \eqref{eq:angelinajolie} with vanishing scalar potential, $V(\phi)=0$, such that in flat space domain wall coordinates \eqref{eq:CH32}  metric and scalar field have the following properties:
\begin{enumerate}
 \item for $r\to\infty$ the metric asymptotes to the null orbifold \eqref{eq:CH36} and the scalar field approaches zero
 \item for $r\to-\infty$ the metric asymptotes to the null orbifold, with a possible rescaling of the first term as in \eqref{eq:CH39} (with some positive $\lambda$) and the scalar field approaches a constant (that can be zero)
 \item for finite values of $r$ the metric function $A(r)$ and the scalar field $\phi(r)$ are bounded real functions; moreover, $A(r)$ is at least $C^2$ and $\phi(r)$ at least $C^1$
\end{enumerate}

By {\em UV (IR)}, we always mean the limits $\rho,r\to\infty$ ($\rho,r\to-\infty$) in the domain wall coordinates referred to above. 

The Virasoro central charges appearing in domain wall solutions of AdS$_3$-Einstein-dilaton gravity are, therefore, denoted by $c^{\textrm{\tiny UV}}$ at the UV boundary and by $c^{\textrm{\tiny IR}}$ at the IR boundary. The Zamolodchikov $c$-theorem implies
\eq{
c^{\textrm{\tiny IR}}\leq c^{\textrm{\tiny UV}}\,.
}{eq:zamo}

Finally, note that the definitions above imply that domain walls always connect UV and IR fixed points, i.e., cases where we do not have an IR fixed point, such as the one discussed at the end of section \ref{sec:4.3} (positive $\alpha$), are excluded by our definitions of proper domain walls.

\subsection{Correspondence theorem}\label{sec:5.1}

Equipped with the definitions of section \ref{sec:5.0}, we can now formulate our first theorem. It allows to translate any proper AdS$_3$ domain wall solution into a corresponding proper flat domain wall solution.

\begin{thm}[AdS$\boldsymbol{_3}$/flat space domain wall correspondence]\label{thm:1}
 Given a proper domain wall solution of AdS$_3$-Einstein-dilaton gravity, there is a corresponding proper flat-space domain wall solution with the following properties:
 \begin{itemize}
  \item[(I.)] In the UV, the flat space asymptotic symmetries generate a $\mathfrak{bms}_3$ algebra with central charge $c_\textrm{M}^{\textrm{\tiny UV}}=\frac{3}{\GN}$.
  \item[(II.)] In the IR, the flat space asymptotic symmetries generate a $\mathfrak{bms}_3$ algebra with a central charge $c_\textrm{M}^{\textrm{\tiny IR}}$ that in general differs from $c_\textrm{M}^{\textrm{\tiny UV}}$.
 \end{itemize}
\end{thm}

\textit{Proof.} Start with some scalar field $\phi(\rho)$ that generates a proper domain wall solution of AdS$_3$-Einstein-dilaton gravity and define $\phi(r)$ to be the scalar field of the corresponding proper flat space domain wall. Since by assumption, the AdS$_3$ domain wall is proper, also the flat space domain wall is proper, meaning there are no singularities at finite values of $r$. Therefore, we need to consider only the UV and IR limits of the scalar field and the metric. By definition we have the expansions $\phi(r\to\infty)=o(1)$ and $\phi(r\to-\infty)=\phi_1 + o(1)$. Since $\phi$ is differentiable we have $\phi'(r\to\infty)=o(1/r)=\phi'(r\to-\infty)$. Integrating the equation of motion \eqref{eq:CH34a} yields $A(r\to\infty)=A_1 r + A_0 + o(1)$ and $A(r\to-\infty)=A_2 r + A_3 + o(1)$. The quantities $A_i$ are integration constants; only two of them can be chosen independently. Without loss of generality, we fix $A_1=1$ and set $A_0=-r_0$, thereby recovering the expansion \eqref{eq:CH35a}. According to the discussion at the beginning of section \ref{sec:3.1}, we then recover the $\mathfrak{bms}_3$ algebra as asymptotic symmetry algebra in the UV with the usual central charge $c_\textrm{M}=3/\GN$. Similarly, we recover a $\mathfrak{bms}_3$ algebra as asymptotic symmetry algebra in the IR, but with a value of the central charge that depends on $A_2$, according to the discussion in the second half of section \ref{sec:3.1}. $\square$ 

The correspondence theorem \ref{thm:1} could be extended to non-proper domain walls (those with no BMS$_3$ fixed point in the IR and instead terminate in a naked singularity), but we refrain from doing so. 

In the final subsection, we prove two additional theorems and start by addressing the paramount issue of monotonicity of the flat space domain wall $c$-function.

\subsection{Monotonicity theorem and central charge ratio equivalence}\label{sec:5.2}

\begin{thm}[Monotonicity of $\boldsymbol{c}$-function]
\label{thm:2}
The $c$-function associated with any flat space domain wall solution obtained through the correspondence theorem \ref{thm:1} is a monotonically decreasing function when flowing from the UV to the IR. 
\end{thm}

\textit{Proof.} Since the scalar field is real, bounded and $C^1$ the metric function $A(r)$ must obey the concavity inequality \eqref{eq:CH41} for all values of $r$. Denoting some fiducial radius as $r_{\textrm{\tiny UV}}$ and another, smaller, fiducial radius as $r_{\textrm{\tiny IR}}<r_{\textrm{\tiny UV}}$, integrating the concavity condition $A''(r)\leq 0$ from $r_{\textrm{\tiny IR}}$ to $r_{\textrm{\tiny UV}}$ implies
\eq{
A'(r_{\textrm{\tiny UV}})\leq A'(r_{\textrm{\tiny IR}})\,.
}{eq:p1}
Inserting this inequality into the definition of the $c$-function \eqref{eq:c},
\eq{
c_{\textrm{\tiny dw}}(r_{\textrm{\tiny UV}})= \frac{c_{\textrm{M}}^{\textrm{\tiny UV}}}{A'(r_{\textrm{\tiny UV}})} \geq \frac{c_{\textrm{M}}^{\textrm{\tiny UV}}}{A'(r_{\textrm{\tiny IR}})} = c_{\textrm{\tiny dw}}(r_{\textrm{\tiny IR}})
}{eq:p2}
establishes that the $c$-function is a monotonically decreasing function when flowing from the UV to the IR, $c_{\textrm{\tiny dw}}(r_{\textrm{\tiny UV}})\geq c_{\textrm{\tiny dw}}(r_{\textrm{\tiny IR}})$.
$\square$

\bigskip

The theorem \ref{thm:2} shows that \eqref{eq:c} is, indeed, a BMS$_3$ $c$-function for proper flat space domain wall solutions. Note that one can consider theorem \ref{thm:2} to be a consequence of bulk unitarity; indeed, if we drop the assumption of the scalar field being real and allow for a purely imaginary scalar field, we can circumvent theorem \ref{thm:2}; the price for this is effectively a switched sign in the kinetic term of the scalar field, which violates bulk unitarity.

We can be more quantitative and combine both theorems to show that the ratio between IR and UV Virasoro central charges is equivalent to the corresponding ratio of $\mathfrak{bms}_3$ central charges.
\eq{
1\leq \frac{c^{\textrm{\tiny UV}}}{c^{\textrm{\tiny IR}}} = \frac{c_\textrm{M}^{\textrm{\tiny UV}}}{c_\textrm{M}^{\textrm{\tiny IR}}} \geq 1
}{eq:thm3}
The first inequality follows from Zamolodchikov's $c$-theorem \cite{Zamolodchikov:1986gt}. The last inequality is the statement of theorem \ref{thm:2} that we just proved. What remains to be shown is the equality in the middle. This central charge ratio equivalence is guaranteed by the third theorem.

\begin{thm}[CFT/CCFT central charge ratio equivalence]\label{thm:3} Given the assumptions of theorem \ref{thm:1}, the ratio of UV/IR central charges obeys the equality in \eqref{eq:thm3}.
\end{thm}

\textit{Proof.} For proper flat space domain walls, the UV/IR ratio of $\mathfrak{bms}_3$ central charges is given by
\eq{
\frac{c_\textrm{M}^{\textrm{\tiny UV}}}{c_\textrm{M}^{\textrm{\tiny IR}}} = \frac{A'(r\to-\infty)}{A'(r\to\infty)} = A_2 \geq 1\,.
}{eq:p3}
The equalities follow from the proof of theorem \ref{thm:1} (and the discussion in section \ref{sec:3.1}). The quantity $A_2$ was also defined in the proof of theorem \ref{thm:1}. The inequality in \eqref{eq:p3} follows from theorem \ref{thm:2}. For proper AdS$_3$ domain walls, the UV/IR ratio of Virasoro central charges is given by the ratio of UV/IR AdS radii. The AdS radii follow from the UV and IR behavior of the function $A(\rho)$ appearing in domain wall coordinates \eqref{eq:CH8}. Since by assumption we set the AdS radius to unity in the UV, we must have the expansion $A(\rho\to\infty)=\rho + \tilde A_0 + o(1)$. Without loss of generality, we equate $\tilde A_0=A_0$ by a constant shift of $\rho$. In the IR we have the expansion $A(\rho\to-\infty)=\tilde A_2 \rho + \tilde A_3 + o(1)$. Therefore, the UV/IR ratio of Virasoro central charges is given by  
\eq{
\frac{c^{\textrm{\tiny UV}}}{c^{\textrm{\tiny IR}}} = \frac{A'(\rho\to-\infty)}{A'(\rho\to\infty)} = \tilde A_2 \geq 1\,.
}{eq:p4}
What remains to be shown is $A_2=\tilde A_2$. Since $A(r)$ and $A(\rho)$ have the same leading and next-to-leading order terms in the UV, it is sufficient to show that both of them obey the same second-order differential equation $A''=-\frac12\,(\phi')^2$. For flat space domain walls, this follows from \eqref{eq:CH34a}. For AdS$_3$ domain walls, this follows from differentiating the left equation \eqref{eq:CH10} with respect to $\rho$ and, using the chain rule, insert the right equation \eqref{eq:CH10} on the right-hand side of the left equation, viz., $\extd^2 A/\extd\rho^2=-\frac12\,\extd W/\extd\phi \cdot \extd\phi/\extd\rho=-\frac12\,(\extd\phi/\extd\rho)^2$. Since $A(\rho)$ and $A(r)$ obey the same second-order differential equation and have the same linear and constant terms in the UV, these two functions must coincide for all values of the radial coordinates. This implies, in particular, $\tilde A_2=A_2$.
$\square$

\bigskip

In conclusion, the three theorems proven in this section provide Carrollian $c$-functions \eqref{eq:cdef} of domain wall solutions \eqref{eq:CH32} to 3d Einstein-dilaton gravity that describe flat space holographic RG flows from a Carrollian UV fixed point to a Carrollian IR fixed point. Bulk unitarity guarantees the monotonicity of our domain wall $c$-functions. Moreover, to every AdS$_3$ domain wall solution (reviewed in section \ref{sec:2.3}), there is a corresponding flat space domain wall solution (discussed in section \ref{sec:4}) with the same radial profile and the same UV/IR-ratios of central charges. The principal difference is that AdS$_3$ domain walls require a scalar field potential for support, whereas flat space domain walls demand vanishing potential.

The drawback of our $c$-functions \eqref{eq:cdef} is that we need some bulk dual, which may not always be available. Thus, it would be satisfying to have an intrinsic construction for a Carrollian $c$-function without recourse to holography, either along the lines of Zamolodchikov's original design \cite{Zamolodchikov:1986gt} or the CH construction reviewed in sections \ref{sec:2.1}. 

We tentatively follow the latter path in our final section, inspired by the relation between the CH $c$-function and QNEC$_2$ recapitulated in section \ref{sec:2.2}. We are guided by the considerations of sections \ref{sec:2.1}, \ref{sec:2.2}, \ref{sec:3.2}, and \ref{sec:3.3}. 

\section{Tentative proposal for Casini--Huerta-inspired Carrollian \texorpdfstring{$\boldsymbol{c}$}{c}-function}\label{sec:6}

Without further ado, here is our tentative proposal for the Carrollian $c$-functions in CCFT$_2$:
\eq{
\boxed{
\phantom{\Big(}
c_{\textrm{L}}(\Delta u,\,\Delta\varphi) := 6 \Delta\varphi\,S_L^\prime\qquad c_{\textrm{M}}(\Delta u,\,\Delta\varphi) := 6\Delta\varphi\,\dot S_M
\phantom{\Big)}}
}{eq:CH28}
Prime denotes $\varphi$-derivatives and dot $u$-derivatives.

A sanity check that our proposal is not ruled out immediately is to consider the special case $c_{\textrm{M}}=0$, $c_{\textrm{L}}\neq 0$ corresponding to a chiral half of a CFT$_2$. In this case, the identity
\eq{
\frac{1}{6\Delta\varphi}\,c_{\textrm{L}}^\prime = S_L'' + \frac{6}{c_{\textrm{L}}}\,S^{\prime\,2}_L
}{eq:CH29}
recovers the expected QNEC$_2$ combination, see the discussion in section \ref{sec:2.2}. Thus, for $c_{\textrm{M}}=0$, $c_{\textrm{L}}\neq 0$, we recover the CH $c$-function for a chiral half of a CFT$_2$.

Another consistency check is that our definitions are independent of the UV cut-offs, as expected on physical grounds.

Finally, even in the more interesting case $c_{\textrm{L}}=0$, $c_{\textrm{M}}\neq 0$, the $c$-function $c_{\textrm{M}}$ reproduces the quantum energy combination of terms \eqref{eq:CH27}.
\eq{
\frac{1}{6\Delta\varphi}\,c_{\textrm{M}}^\prime = \dot S_M^\prime + \frac{6}{c_{\textrm{M}}}\,\dot S^2_M
}{eq:CH30}
Thus, if the $c$-function $c_{\textrm{M}}$ is monotonic, it implies the quantum energy condition \eqref{eq:CH27} for the ground state. 

The arguments above are neither proof of the quantum energy conditions nor proof that $c_{\textrm{M}}$ is a $c$-function; they merely show that our putative $c$-function in \eqref{eq:CH28} is consistent with the CCFT$_2$ quantum energy conditions \cite{Grumiller:2019xna}. We leave applications and scrutiny of our proposal \eqref{eq:CH28} to future work.

\acknowledgments

We thank Luis Apolo, Arjun Bagchi, Rudranil Basu, Jacqueline Caminiti, Rob Myers, and Wei Song for discussions on flat space holographic EE.

This work was supported by the Austrian Science Fund (FWF), projects P~32581, P~33789, and P~36619. Some of our results were presented in March 2021 at the virtual workshop ``Flat Asymptotia'' organized by Aritra Banerjee, Sudip Ghosh, Slava Lysov, and Yasha Neiman. This research was supported in part by Perimeter Institute for Theoretical Physics. Research at Perimeter Institute is supported by the Government of Canada through the Department of Innovation, Science and Economic Development and by the Province of Ontario through the Ministry of Colleges and Universities. The final part of this research was conducted while DG was visiting the Okinawa Institute of Science and Technology (OIST) through the Theoretical Sciences Visiting Program (TSVP) in July/August 2023.

\section*{Note added}

After posting our paper on the {\tt arXiv} we were apprised that two of our main results and one of our theorems were published in \cite{Fareghbal:2015bxd}.

In particular, the flat space domain wall geometries \eqref{eq:CH32} correspond to their (4.11) upon redefining our radial coordinate as $\extd r\to e^{-A(r)}\extd r$, the domain wall $c$-function \eqref{eq:cdef} essentially corresponds to their (4.16), and theorem \ref{thm:2} corresponds to their statement after (4.17).

New results in our work not contained in \cite{Fareghbal:2015bxd} include the proof that flat space domain walls are solutions to Einstein-dilaton gravity without scalar field potential, a discussion of their curvature invariants, the flat space holographic RG flow example, theorems \ref{thm:1} and \ref{thm:3}, and the tentative proposal for the CH $c$-function.

%\bibliographystyle{fullsort}
%\bibliography{review}

\providecommand{\href}[2]{#2}\begingroup\raggedright\endgroup

\end{document}